\documentclass{aastex63}

\usepackage{color}
\usepackage{graphicx}
\usepackage{amsmath}

\received{}
\revised{}
\accepted{}
\submitjournal{ApJ}

\shorttitle{An ALMA-ACA CO Survey in the SMC North II}
\shortauthors{Ohno et al.}

\begin{document}

\title{An Unbiased CO Survey Toward the Northern Region of the Small Magellanic Cloud with the Atacama Compact Array. II. CO Cloud Catalog}

\correspondingauthor{Takahiro Ohno, Kazuki Tokuda}
\email{t.ohno4a.phys@gmail.com, tokuda.kazuki.369@m.kyushu-u.ac.jp}

\author{Takahiro Ohno}
\affiliation{Department of Physics, Nagoya University, Chikusa-ku, Nagoya 464-8602, Japan}

\author[0000-0002-2062-1600]{Kazuki Tokuda}
\affiliation{Department of Earth and Planetary Sciences, Faculty of Science, Kyushu University, Nishi-ku, Fukuoka 819-0395, Japan}
\affiliation{National Astronomical Observatory of Japan, National Institutes of Natural Sciences, 2-21-1 Osawa, Mitaka, Tokyo 181-8588, Japan}
\affiliation{Department of Physics, Graduate School of Science, Osaka Metropolitan University, 1-1 Gakuen-cho, Naka-ku, Sakai, Osaka 599-8531, Japan}

\author{Ayu Konishi}
\affiliation{Department of Physics, Graduate School of Science, Osaka Metropolitan University, 1-1 Gakuen-cho, Naka-ku, Sakai, Osaka 599-8531, Japan}

\author{Takeru Matsumoto}
\affiliation{Department of Physics, Graduate School of Science, Osaka Metropolitan University, 1-1 Gakuen-cho, Naka-ku, Sakai, Osaka 599-8531, Japan}

\author{Marta Sewi{\l}o}
\affiliation{CRESST II and Exoplanets and Stellar Astrophysics Laboratory, NASA Goddard Space Flight Center, Greenbelt, MD 20771, USA}
\affiliation{Department of Astronomy, University of Maryland, College Park, MD 20742, USA}

\author{Hiroshi Kondo}
\affiliation{Department of Physics, Graduate School of Science, Osaka Metropolitan University, 1-1 Gakuen-cho, Naka-ku, Sakai, Osaka 599-8531, Japan}

\author{Hidetoshi Sano}
\affiliation{Faculty of Engineering, Gifu University, 1-1 Yanagido, Gifu 501-1193, Japan}

\author{Kisetsu Tsuge}
\affiliation{Dr. Karl Remeis Observatory and ECAP, Universit\"{a}t Erlangen-N\"{u}rnberg, Sternwartstrasse 7, D-96049, Bamberg, Germany}

\author[0000-0001-6149-1278]{Sarolta Zahorecz}
\affiliation{National Astronomical Observatory of Japan, National Institutes of Natural Sciences, 2-21-1 Osawa, Mitaka, Tokyo 181-8588, Japan}
\affiliation{Department of Physics, Graduate School of Science, Osaka Metropolitan University, 1-1 Gakuen-cho, Naka-ku, Sakai, Osaka 599-8531, Japan}

\author{Nao Goto}
\affiliation{Department of Physics, Graduate School of Science, Osaka Metropolitan University, 1-1 Gakuen-cho, Naka-ku, Sakai, Osaka 599-8531, Japan}

\author[0000-0001-8901-7287]{Naslim Neelamkodan}
\affiliation{Department of physics, United Arab Emirates University, Al-Ain, 15551, UAE}

\author[0000-0002-7759-0585]{Tony Wong}
\affiliation{Department of Astronomy, University of Illinois, Urbana, IL 61801, USA}

\author{Hajime Fukushima}
\affiliation{Center for Computational Sciences, University of Tsukuba, Ten-nodai, 1-1-1 Tsukuba, Ibaraki 305-8577, Japan}

\author[0000-0002-4124-797X]{Tatsuya Takekoshi}
\affiliation{Kitami Institute of Technology, 165 Koen-cho, Kitami,
Hokkaido 090-8507, Japan}

\author{Kazuyuki Muraoka}
\affiliation{Department of Physics, Graduate School of Science, Osaka Metropolitan University, 1-1 Gakuen-cho, Naka-ku, Sakai, Osaka 599-8531, Japan}

\author{Akiko Kawamura}
\affiliation{National Astronomical Observatory of Japan, National Institutes of Natural Science, 2-21-1 Osawa, Mitaka, Tokyo 181-8588, Japan}

\author{Kengo Tachihara}
\affiliation{Department of Physics, Nagoya University, Chikusa-ku, Nagoya 464-8602, Japan}

\author{Yasuo Fukui}
\affiliation{Department of Physics, Nagoya University, Chikusa-ku, Nagoya 464-8602, Japan}

\author[0000-0001-7826-3837]{Toshikazu Onishi}
\affiliation{Department of Physics, Graduate School of Science, Osaka Metropolitan University, 1-1 Gakuen-cho, Naka-ku, Sakai, Osaka 599-8531, Japan}

\begin{abstract}

The nature of molecular clouds and their statistical behavior in subsolar metallicity environments are not fully explored yet. We analyzed data from an unbiased CO($J$ = 2--1) survey at the spatial resolution of $\sim$2\,pc in the northern region of the Small Magellanic Cloud with the Atacama Compact Array to characterize the CO cloud properties. A cloud-decomposition analysis identified 426 spatially/velocity-independent CO clouds and their substructures. Based on the cross-matching with known infrared catalogs by Spitzer and Herschel, more than 90\% CO clouds show spatial correlations with point sources. We investigated the basic properties of the CO clouds and found that the radius--velocity linewidth ($R$--$\sigma_{v}$) relation follows the Milky Way-like power-low exponent, but the intercept is $\sim$1.5 times lower than that in the Milky Way.
The mass functions ($dN/dM$) of the CO luminosity and virial mass are characterized by an exponent of $\sim$1.7, which is consistent with previously reported values in the Large Magellanic Cloud and in the Milky Way.

\end{abstract}


\section{Introduction} \label{sec:intro}
Molecular clouds are the densest phase of the interstellar medium (ISM) and the fundamental cradles for star formation. Although the primary ingredient of molecular gas is hydrogen molecules, its direct measurement is almost impossible due to the lack of a suitable transition under the typical condition of molecular clouds. Alternative tracers, such as low-$J$ transitions of CO and thermal dust emission, have been used to reveal the molecular cloud distribution and properties. 
In the past few decades, observations with the millimeter-wave facilities improved our understanding of the nature of molecular clouds in the solar neighborhood and the Galactic Plane (see the review by \citealt{Heyer15}). Molecular clouds traced by CO observations in the Milky Way (MW) follow a standard size--linewidth relation (see \citealt{Solomon87}), and the CO luminosity and the mass are well correlated with each other, suggesting that the molecular clouds are in virial equilibrium as a whole \citep[e.g.,][]{Larson81}. 

These observations have not been limited to the MW, but extended to some of the other galaxies in the Local Group (e.g., \citealt{Cohen88,Rubio91,Fukui99,Engargiola03,Nieten06}; see the review by \citealt{Fukui10}), providing information on the statistical properties of molecular clouds. However, some of the extreme conditions are not fully explored yet in high spatial resolution due to observational difficulties. Among them, low-metallicity environments are a good frontier for understanding the star formation in the early universe. The MW observations indicate that the metallicity decreases with increasing galactocentric radius, down to subsolar values \citep{Fer17}, and thus the outermost part in the Galactic disk is suitable for studying this aspect, and some surveys confirmed the presence of CO clouds \citep[e.g.,][]{Dame11b,Izumi14,Matsuo17}. Unfortunately, distance ambiguities and contamination in the same line of sight in the Galactic plane are always problems for us when a uniform sample is to be obtained and statistical analyses are to be performed.

In this regard, the Small Magellanic Cloud (SMC), with a metallicity of $\sim$0.2 $Z_{\odot}$ \citep{Russell92,Rolleston99,Pagel03}, is above the Galactic plane, providing a suitable condition for observing the entire galaxy and understanding the behavior of the CO cloud through its favorable spatial extension and proximity ($\sim$62\,kpc, \citealt{Graczyk20}). The metallicity is close to that in the early universe, showing active star formation \citep{Pei99}, and thus, it is desirable to obtain the fundamental parameter from spatially resolved observations, such as the CO-to-H$_2$ conversion factor (hereafter, $X_{\rm CO}$), to understand the gas properties of more remote galaxies. \cite{Rubio91} and \cite{Mizuno01} performed large-scale CO surveys with an angular resolution of 8\farcm8 or 2\farcm6, corresponding to 160--45\,pc. 
They derived an $X_{\rm CO}$ in the SMC of (2.5--6)\,$\times$10$^{21}$\,cm$^{-2}$\,(K\,km\,s$^{-1}$)$^{-1}$, which is 10--20 times higher than the canonical Galactic value of $\sim$2\,$\times$10$^{20}$\,cm$^{-2}$\,(K\,km\,s$^{-1}$)$^{-1}$ \citep[e.g.,][]{Dame01,Bolatto13}, by comparing the CO luminosity and the dynamical (virial) mass. However, these studies also suggested that the beam-filling factor of CO clouds in the SMC is smaller than that of the Galactic molecular cloud, and the large-beam measurements introduce large uncertainties in the analysis, even if the virial equilibrium assumption is reasonable. 
Several theoretical studies proposed that the $X_{\rm CO}$ factor depends on metallicity with a power-low index of $-$(0.5--0.8) \citep[e.g.,][]{Feldmann12}, indicating that the sub-solar metallicity condition does not significantly change the $X_{\rm CO}$ factor from that of the MW value. Smaller beam size measurements indeed yielded lower values than the above surveys, although the results are based on only partial observations compared to the entire galaxy \citep[e.g.,][]{Bolatto03,Muraoka17,Jameson18,ONeill22}. The fundamental properties of molecular clouds, such as the mass function and size--linewidth relation, are not necessarily sufficiently obtained by compiling a statistically large sample. \cite{Saldano22} recently presented the SMC CO(2--1) survey at a resolution of 9\,pc using the Atacama Pathfinder Experiment (APEX) telescope and obtained basic properties of molecular clouds across the galaxy. Nevertheless, higher-sensitivity and higher-resolution data are still needed for a complete census, including low-mass and infrared-quiescent clouds, whose CO intensities are generally weak.

The Atacama Large Millimeter/submillimeter Array (ALMA) has the potential to perform a high-resolution unbiased survey of a relatively small galaxy. Especially, the Atacama Compact Array (ACA), known as the Morita array, is not only sensitive to a low-spatial frequency component, which is probably advantageous for capturing an extended CO cloud, but also has the advantage of a wider field of view than the 12\,m array (the ALMA Main array), making it a powerful survey instrument. In our companion paper of \cite{Tokuda21} (hereafter Paper~I), we described the ALMA archival CO survey covering $\sim$0.26\,deg$^2$ in the northern SMC, assessed the data quality, and provided the initial results of the data analysis. The present paper includes a detailed discussion of the CO cloud decomposition and a statistical analysis to understand the role of CO as molecular cloud tracer. Section~\ref{sec:data} summarizes the CO data that we use in this manuscript, and then we present the identification method of CO clouds and their characterization in Section~\ref{sec:results}. The discussions and summary are presented in Sections~\ref{sec:dis} and \ref{sec:sum}, respectively.

\section{The data} \label{sec:data}
This study uses the ALMA archival CO data in the SMC north region (2017.A.00054.S), which was proposed by the ALMA observatory as one of the six filler programs\footnote{https://almascience.nao.ac.jp/news/alma-announces-aca-observatory-filler-programs-for-cycle-6} for the ACA stand-alone mode. Because Paper~I described the survey setting and data reduction in detail, we briefly summarize the data quality here. The available data set includes the CO($J$ = 2--1) and CO($J$ = 1--0) lines, and 1.3/2.6\,mm continuum data with a field coverage of $\sim$0.26\,deg$^{^2}$.
The angular resolution and sensitivity of CO($J$ = 2--1) are 6\farcs9$\times$6\farcs6 ($\sim$2\,pc) and $\sim$0.06\,K, respectively. The Cube data with a velocity-channel width of 0.5\,km\,s$^{-1}$ were used throughout the analysis in this paper. The resultant detection limit in the CO($J$ = 2--1) luminosity is $\sim$1.0\,K\,km\,s$^{-1}$\,pc$^{2}$ (Paper~I). Note that the angular resolution and sensitivity of the CO($J$ = 2--1) data are two and four times better than those in the CO($J$ = 1--0) data, respectively, and thus we mainly use the former data in this study.

\section{Results} \label{sec:results}

\subsection{Cloud Decomposition}\label{sec:dendro}

Interstellar molecular clouds generally have hierarchical, complex structures composed of diffuse gas, dense filaments, and cores \citep[e.g.,][]{Lada92}. The complexity in nature makes it difficult for us to determine clear boundaries of each subcomponent; nevertheless, some decomposition analyses, which have been developed in the last decades \citep[e.g.,][]{Williams94, Rosolowsky06, Rosolowsky08}, are still powerful tools for characterizing cloud properties and their statistical nature, such as the size--linewidth relation and mass function. As described in Paper~I, the CO molecular clouds in the SMC are spatially more compact than those in the MW, and the outer boundaries are relatively easy to define. On the other hand, larger clouds in the observed field have multiple local peaks inside, requesting a hierarchical characterization of the structure with different intensity levels. 
\cite{Bolatto13} suggested that the properties of the outer and inner regions of molecular clouds are somewhat different in low-metallicity environments, such as the SMC. Therefore, it is useful to treat the large outer and small inner structures separately. The dendrogram algorithm, \texttt{astrodendro} \citep{Rosolowsky08,Shetty12,Colombo15} is one of the best options to meet our requirements (see also the comparison of different cloud-decomposition methods by \citealt{Li20}). Several studies \citep{Wong17,Naslim18,Nayak18,Wong19} applied the same scheme to ALMA CO data of molecular clouds in the Large Magellanic Cloud (LMC) at an angular resolution of $\sim$1\,pc.  
The CPROPS method of \cite{Rosolowsky06} is also promising, but there are limitations in decomposing physically reasonable objects in highly crowded and low-contrast environments \citep{Colombo14}. A patchwork-like separation using CLUMPFIND \citep{Williams94} enables us to estimate the total flux of discrete objects, but large and small structures cannot be treated separately.

As input data, we used postprocessed CO cube, moment-masked data (see \citealt{Dame11a}) whose emission-free pixels were set at zero value judging from a smoothed data cube whose signal-to-noise ratio is higher than that of the raw data (see also the detailed description in Section~2 in Paper~I). The \texttt{astrodendro} algorithm has three input parameters, \texttt{min\_value}, \texttt{min\_npix}, and \texttt{min\_delta}. The first argument is the minimum-intensity value to consider in the cube data. 
Because most of the noise-component pixels have already been eliminated by the masking analysis, we decided to consider emissions that were as weak as possible by setting \texttt{min\_value} to 0\,K. This zero-level setting minimizes the truncation effect of weak emission and does not account for unreliable weak peaks. The combination with the other two parameters described below resulted in a significant cloud identification with a lowest peak intensity of 0.35\,K ($\gtrsim$5$\sigma$) among all entities. The second parameter, \texttt{min\_pix}, is the minimum number of voxels that have significant emission in the three-dimensional ($x$,$y$,$v$) axis needed to be connected as a single component. We set this value of 38 equal to the voxel number of at least a single-beam element in $XY$ space and three pixels in the velocity direction. These two parameters are well defined by the setting of the observation, and thus, we treat them as fixed values, while the last parameter, \texttt{min\_delta}, can be chosen arbitrarily. The value is a threshold for entities in close proximity to be considered as independent components. Our fiducial value of \texttt{min\_delta} is 0.18\,K, corresponding to a noise level of $\sim$3$\sigma$ for the data set. The number of identified structures and statistical results does not change significantly even if this value is changed by a factor of several from the fiducial value. Although we decomposed the cloud and discussed the data using the fixed fiducial value, the parameter dependence is further discussed in the Appendix.

\begin{figure}[htbp]
\centering
\includegraphics[width=180mm]{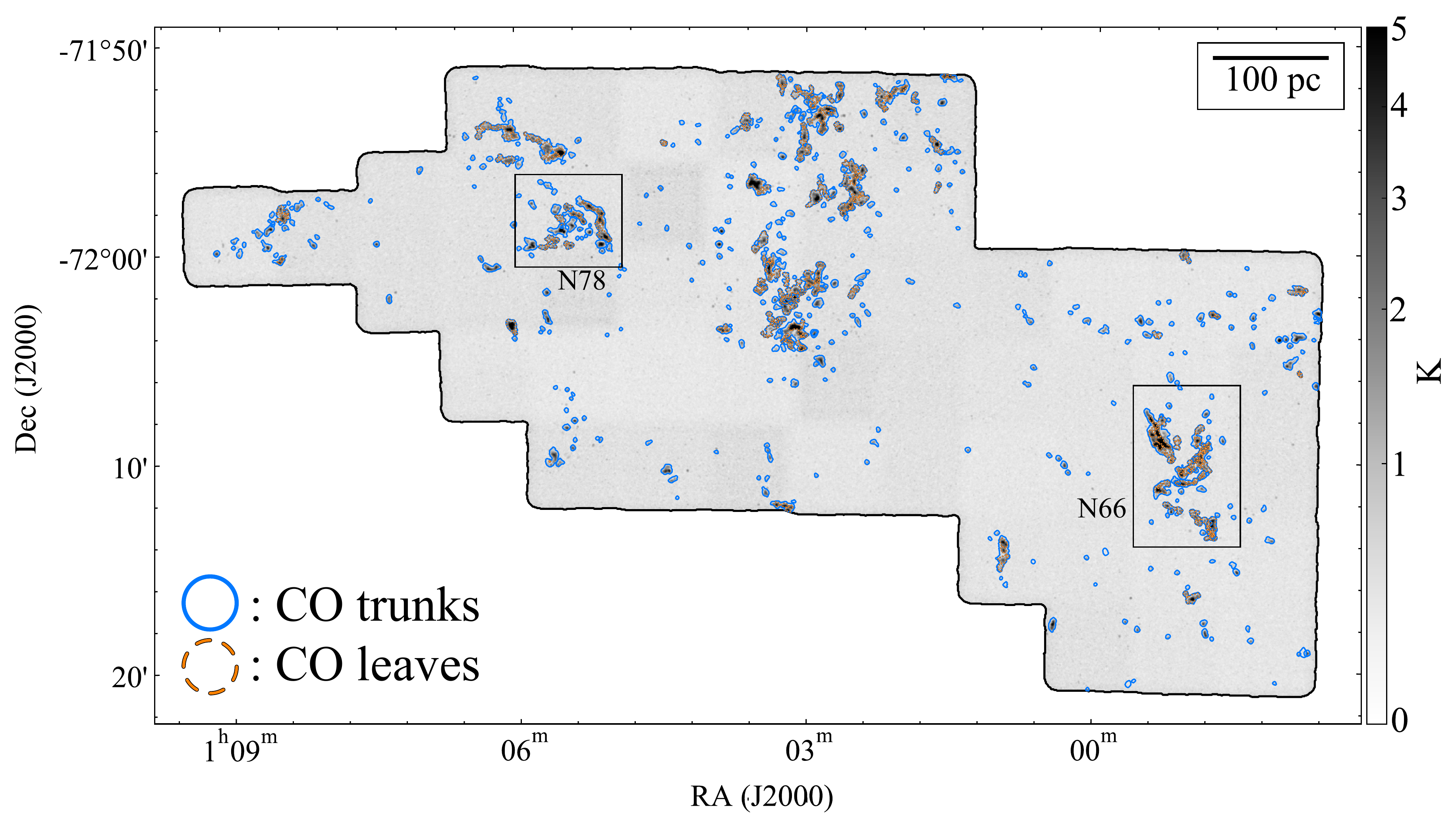}
\caption{Distributions of the identified structures on the CO($J$ = 2--1) map of the SMC northern region. The grayscale image shows the peak brightness temperature map in CO($J$ = 2--1) obtained with the ACA. The solid cyan and dashed orange contours denote the boundaries of the CO trunks and laves, respectively, that were identified by the \texttt{astrodendro} algorithm.
The two rectangles show the areas displayed in Figure~\ref{fig:dendrozoom}.}  
\label{fig:dendro_north}
\end{figure}

\begin{figure}[htbp]
\centering
\includegraphics[width=150mm]{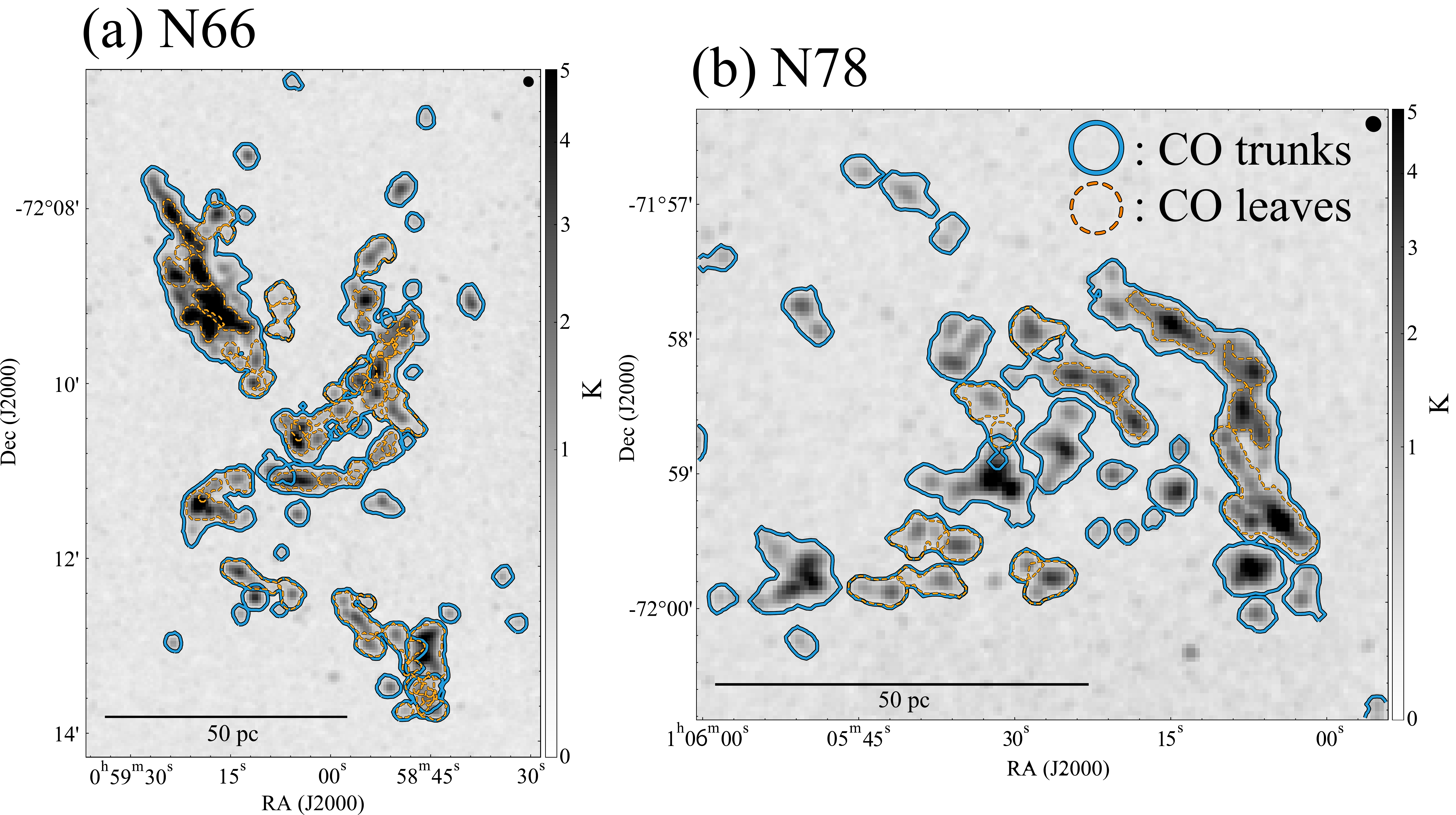}
\caption{Zoomed-in views of the CO trunks and leaves on the CO($J$ = 2--1) map toward the N66 and N78 regions. (a) The grayscale image, the solid cyan and dashed orange contours show the peak brightness temperature map, the CO trunk and leaf boundaries, respectively, toward the N66 region. 
The black ellipse in the upper right corner is the angular resolution of the CO($J$ = 2--1) data, 6\farcs9$\times$6\farcs6. (b) Same as (a), but for the N78 region.}
\label{fig:dendrozoom}
\end{figure}

We performed the \texttt{astrodendro} analysis and identified 426 structures, called trunks, which are the largest continuous structures. Of these, 361 trunks do not contain internal structures and are categorized as single CO trunks, which are spatially/velocity-independent entities of the surroundings. In addition, 65 trunks contain internal structures (referred to as CO leaves) for a total of 257 internal leaves. We refer to the 426 trunks and 257 internal leaves as CO trunks and CO leaves, respectively.
Figure~\ref{fig:dendro_north} illustrates the boundary of individual sources of the two categories on the CO map. Figure~\ref{fig:dendrozoom} shows two examples of zoomed-in views toward the N66 and N78 regions to demonstrate how the identified structures are distributed in the two large systems. The CO trunk boundaries are determined by an isosurface close to the minimum-intensity contour level in the data cube, providing a fairly robust identification against the input parameter dependence. The 2D projected map sometimes shows overlapping boundaries, but they are independent entities in velocity space. The dependence of \texttt{min\_delta} is somewhat more sensitive in the CO leaves than in the trunks. Nevertheless, the CO leaf boundary seems to reasonably trace local peaks on the CO map (see Figure~\ref{fig:dendrozoom}).

The \texttt{astrodendro} analysis outputs the basic properties of the identified structures, their centroid coordinates in three-dimensional axes ($\overline{x}$, $\overline{y}$, $\overline{v}$), the rms size of the major/minor axes ($\sigma_{\rm maj}$ and $\sigma_{\rm min}$), the rms linewidth $\sigma_{v}$, and the position angle of the major axis (P.A.). Within the isosurface contours of all identified structures, we additionally derived several parameters. The brightness temperature $T_{\rm peak}$ is simply the peak value of the identified voxels. We integrated the flux to obtain the CO($J$ = 2--1) luminosity $L_{\rm CO(2\text{--}1)}$, adopting a distance of 62\,kpc \citep{Graczyk20}. The effective rms size, $\sigma_{r} = \sqrt{\sigma_{\rm maj}\sigma_{\rm min}}$, is multiplied by 1.91, as suggested by \cite{Solomon87}, to derive the observed spherical radius $R_{\rm obs}$, and then we applied the beam-deconvolution scheme, $R_{\rm deconv} = \sqrt{R_{\rm obs}^2 - \theta_{\rm beam}^2}$, where $\theta_{\rm beam}$ is the beam size of the present study. We used an approach to estimate the uncertainties of the cloud properties following the bootstrap method \citep{Rosolowsky06}. We generated 100 realizations to sample the derived parameters. Tables~\ref{table:trunk} and \ref{table:leaf} summarize the properties of some of the identified CO trunks and leaves, respectively, and the full catalogs are available as online material.

\begin{deluxetable}
{lccccccccccc}
\tabletypesize{\scriptsize}
\tablecaption{Physical Properties of the CO Trunks \label{table:trunk}}
\tablewidth{0pt}
\tablehead{
\colhead{id} & \colhead{R.A. (J2000.0)} & \colhead{Dec. (J2000.0)} & \colhead{$\overline{v}$} & \colhead{$\sigma_{v}$} & \colhead{$\delta\sigma_{v}$} & \colhead{$\sigma_{\rm maj}$} & \colhead{$\sigma_{\rm min}$} & \colhead{P.A.} & \colhead{$T_{\rm peak}$} & \colhead{$L_{\rm CO(2\text{--}1)}$} & \colhead{$\delta L_{\rm CO(2\text{--}1)}$}\\
& \colhead{[hms]} & \colhead{[dms]} & \colhead{[km s$^{-1}$]} & \colhead{[km s$^{-1}$]} & \colhead{[km s$^{-1}$]} & \colhead{[arcsec.]} & \colhead{[arcsec.]} & \colhead{[deg.]} & \colhead{[K]} & \colhead{[K km s$^{-1}$ pc$^{2}$]} & \colhead{[K km s$^{-1}$ pc$^{2}$]}\\
(1)&(2)&(3)&(4)&(5)&(6)&(7)&(8)&(9)&(10)&(11)&(12)}
\startdata
0	&$	0	^{\mathrm{h}}	57	^{\mathrm{m}}	46\fs8	$&$	-72	{^\circ}	19	'	04\farcs8	$&	116.1 	&	0.52 	&	0.13 	&	\phn5.5	&	3.4	&$	\phantom{-}\phn94.5	$&	2.5	& \phn41.7	&	0.6\\
1	&$	0	^{\mathrm{h}}	59	^{\mathrm{m}}	52\fs3	$&$	-72	{^\circ}	14	'	24\farcs0	$&	117.0 	&	0.37 	&	0.18 	&	\phn4.3	&	2.7	&$	\phantom{-}110.4	$&	2.8 & \phn21.3	&	1.1\\
2	&$	0	^{\mathrm{h}}	59	^{\mathrm{m}}	33\fs4	$&$	-72	{^\circ}	20	'	34\farcs8	$&	121.2 	&	0.34 	&	0.27 	&	\phn2.5	&	1.7	&$	-156.6	$&	1.1 & \phn\phn3.9	&	1.4\\
3	&$	0	^{\mathrm{h}}	59	^{\mathrm{m}}	37\fs0	$&$	-72	{^\circ}	20	'	42\farcs0	$&	122.1 	&	0.70 	&	0.15 	&	\phn3.4	&	2.6	&$	\phantom{-}\phn71.0	$&	2.9 & \phn31.9	&	0.8\\
4	&$	0	^{\mathrm{h}}	59	^{\mathrm{m}}	58\fs3	$&$	-72	{^\circ}	15	'	57\farcs6	$&	121.7 	&	0.38 	&	0.28 	&	\phn2.1	&	1.9	&$	-142.1	$&	0.4 & \phn\phn1.9	&	0.8\\
5	&$	1	^{\mathrm{h}}	00	^{\mathrm{m}}	37\fs7	$&$	-72	{^\circ}	14	'	56\farcs4	$&	128.1 	&	0.43 	&	0.22 	&	\phn2.4	&	2.2	&$	\phantom{-}114.9	$&	1.4 & \phn\phn6.7	&	0.9\\
6	&$	0	^{\mathrm{h}}	59	^{\mathrm{m}}	04\fs1	$&$	-72	{^\circ}	11	'	02\farcs4	$&	144.3 	&	2.02 	&	0.06 	&	14.4	&	3.5	&$	-175.8	$&	4.4 & 343.0 	&	0.3\\
7	&$	0	^{\mathrm{h}}	58	^{\mathrm{m}}	22\fs6	$&$	-72	{^\circ}	12	'	46\farcs8	$&	144.6 	&	0.55 	&	0.21 	&	\phn2.7	&	2.4	&$	\phantom{-}\phn76.2	$&	2.0 & \phn13.5	&	0.9\\
8	&$	0	^{\mathrm{h}}	58	^{\mathrm{m}}	52\fs6	$&$	-72	{^\circ}	10	'	40\farcs8	$&	145.1 	&	1.55 	&	0.06 	&	\phn5.8	&	3.6	&$	\phantom{-}\phn48.4	$&	1.9 & \phn72.9	&	0.3\\
9	&$	0	^{\mathrm{h}}	58	^{\mathrm{m}}	04\fs6	$&$	-72	{^\circ}	20	'	31\farcs2	$&	145.2 	&	0.39 	&	0.24 	&	\phn2.3	&	1.9	&$	\phantom{-}138.1	$&	0.8 & \phn\phn3.5	&	1.1\\
\hline
\colhead{id} & \colhead{$R_{\rm deconv}$} & \colhead{$\delta R_{\rm deconv}$} & \colhead{$M_{\rm vir}$} & \colhead{$\delta M_{\rm vir}$} & \colhead{$N_{\rm H_2}$} &    \colhead{$M_{\rm CO}$}	& \colhead{$n_{\rm H_2}$} & \colhead{IR source}	& &	& \\
& \colhead{[pc]} & \colhead{[pc]} & \colhead{[$M_{\odot}$]} & \colhead{[$M_{\odot}$]} & \colhead{[10$^{21}$ cm$^{-2}$]} & \colhead{[$M_{\odot}$]}	& \colhead{[10$^{2}$ cm$^{-3}$]} &  &   &	& \\
(13)&(14)&(15)&(16)&(17)&(18)&(19)&(20)&(21)&&& \\
\hline
0 &		2.26	&	0.08 	&	\phn\phn634		&	310	&   \phn3.0	&   \phn751	&   \phn6	&   B	&	&	& \\
1 &		1.66	&	0.11 	&	\phn\phn235		&	239	&   \phn1.9	&   \phn385	&   \phn8	&   A	&	&	& \\
2 &		0.63	&	0.24 	&	\phn\phn\phn74	&	144	&   \phn0.9	&   \phn\phn69	&   27	&   B 	&	&	& \\
3 &		1.38	&	0.10 	&	\phn\phn696		&	301	&   \phn4.3	&   \phn574	&       21	&   A	&	&	& \\
4 &		0.51	&	0.23 	&	\phn\phn\phn78	&	147	&   \phn0.5	&   \phn\phn34	&   24	&   B	&	&	& \\
5 &		0.88	&	0.18 	&	\phn\phn167		&	192	&   \phn1.3	&   \phn120	&   17	&       B	&	&	& \\
6 &		3.98	&	0.04 	&	16871			&	924	&   12.1	    &   6183	&   \phn9	&   A	&	&	& \\
7 &		1.05	&	0.13 	&	\phn\phn332		&	258	&   \phn2.4	&   \phn243	&   20	&       A	&	&	& \\
8 &		2.41	&	0.06 	&	\phn6000		&	467	&   \phn4.3	&   1315	&   \phn9	&   A	&	&	& \\
9 &		0.65	&	0.20 	&	\phn\phn101		&	148	&   \phn0.8	&   \phn\phn62	&   21 	&   B	&	&	& \\
\enddata
\tablecomments{
$\delta$ denotes the errors, and they are derived using the bootstrap method (see the text in Section~\ref{sec:dendro}).
Column density ($N_{\rm H_2}$), CO luminosity masses ($M_{\rm CO}$), and H$_2$ volume density ($n_{\rm H_2}$) assuming $X_{\rm CO}$ = 7.5 $\times$10$^{20}$\,cm$^{-2}$\,(K\,km\,s$^{-1}$)$^{-1}$ and a CO(2--1)/(1--0) ratio of 0.9. A and B represent Spitzer + Herschel YSO candidate and other Spitzer catalog source, respectively (see the text in Section\ref{sec:resultYSO}).} The full catalog is available as online material. 
\end{deluxetable}

\begin{deluxetable}
{lccccccccccc}
\tabletypesize{\scriptsize}
\tablecaption{Physical Properties of the CO Leaves \label{table:leaf}}
\tablewidth{0pt}
\tablehead{
\colhead{id} & \colhead{R.A. (J2000.0)} & \colhead{Dec. (J2000.0)} & \colhead{$\overline{v}$} & \colhead{$\sigma_{v}$} & \colhead{$\delta\sigma_{v}$} & \colhead{$\sigma_{\rm maj}$} & \colhead{$\sigma_{\rm min}$} & \colhead{P.A.} & \colhead{$T_{\rm peak}$} & \colhead{$L_{\rm CO(2\text{--}1)}$} & \colhead{$\delta L_{\rm CO(2\text{--}1)}$}\\
& \colhead{[hms]} & \colhead{[dms]} & \colhead{[km s$^{-1}$]} & \colhead{[km s$^{-1}$]} & \colhead{[km s$^{-1}$]} & \colhead{[arcsec.]} & \colhead{[arcsec.]} & \colhead{[deg.]} & \colhead{[K]} & \colhead{[K km s$^{-1}$ pc$^{2}$]} & \colhead{[K km s$^{-1}$ pc$^{2}$]}\\
(1)&(2)&(3)&(4)&(5)&(6)&(7)&(8)&(9)&(10)&(11)&(12)}
\startdata
0	&$	0	^{\mathrm{h}}	58	^{\mathrm{m}}	56\fs9	$&$	-72	{^\circ}	10	'	55\farcs2	$&140.5&1.30&0.22&2.6&2.0&$-$140.9&0.9&\phn\phn9.6&0.3\\
1	&$	0	^{\mathrm{h}}	59	^{\mathrm{m}}	01\fs0	$&$	-72	{^\circ}	11	'	02\farcs4	$&142.8&0.75&0.25&2.8&1.5&$-$173.0&2.4&\phn21.1&0.3\\
2	&$	0	^{\mathrm{h}}	59	^{\mathrm{m}}	05\fs5	$&$	-72	{^\circ}	11	'	02\farcs4	$&145.3&1.08&0.11&6.5&2.2&$\phantom{-}$176.3&4.4&142.3&0.3\\
3	&$	0	^{\mathrm{h}}	59	^{\mathrm{m}}	14\fs6	$&$	-72	{^\circ}	11	'	06\farcs0	$&146.4&1.45&0.13&3.3&2.5&$\phantom{-}$155.2&1.9&\phn38.6&0.4\\
4	&$	0	^{\mathrm{h}}	58	^{\mathrm{m}}	53\fs3	$&$	-72	{^\circ}	10	'	44\farcs4	$&144.0&0.47&0.33&3.5&1.7&$\phantom{-}$\phn46.1&1.9&\phn11.9&0.4\\
5	&$	0	^{\mathrm{h}}	58	^{\mathrm{m}}	57\fs1	$&$	-72	{^\circ}	11	'	02\farcs4	$&144.3&0.44&0.29&2.2&1.7&$-$139.0&1.1&\phn\phn5.1&0.7\\
6	&$	0	^{\mathrm{h}}	57	^{\mathrm{m}}	54\fs7	$&$	-72	{^\circ}	01	'	40\farcs8	$&146.7&0.99&0.09&5.0&3.0&$\phantom{-}$171.0&3.2&\phn71.4&0.4\\
7	&$	0	^{\mathrm{h}}	58	^{\mathrm{m}}	52\fs1	$&$	-72	{^\circ}	10	'	37\farcs2	$&146.2&0.62&0.30&3.6&1.5&$\phantom{-}$\phn79.8&1.3&\phn10.0&0.2\\
8	&$	0	^{\mathrm{h}}	58	^{\mathrm{m}}	53\fs5	$&$	-72	{^\circ}	09	'	46\farcs8	$&147.4&0.87&0.17&4.0&1.6&$\phantom{-}$105.1&1.2&\phn\phn9.1&0.3\\
9	&$	0	^{\mathrm{h}}	59	^{\mathrm{m}}	18\fs5	$&$	-72	{^\circ}	11	'	09\farcs6	$&147.2&0.50&0.31&4.9&1.4&$-$137.3&1.7&\phn10.3&0.3\\
\hline
\colhead{id} & \colhead{$R_{\rm deconv}$} & \colhead{$\delta R_{\rm deconv}$} & \colhead{$M_{\rm vir}$} & \colhead{$\delta M_{\rm vir}$} & \colhead{$N_{\rm H_2}$} &    \colhead{$M_{\rm CO}$}	&   \colhead{$n_{\rm H_2}$}  &   &   &	&\\
& \colhead{[pc]} & \colhead{[pc]} & \colhead{[$M_{\odot}$]} & \colhead{[$M_{\odot}$]} & \colhead{[10$^{21}$ cm$^{-2}$]} &	\colhead{[$M_{\odot}$]}	&   \colhead{[10$^{2}$ cm$^{-3}$]}  &	&   &	& \\
(13)&(14)&(15)&(16)&(17)&(18)&(19)&(20)&&&&\\
\hline
0 &0.80&0.15&1406&514&\phn1.9&\phn173&\phn32&&&&\\
1 &0.59&0.14&\phn346&256&\phn4.5&\phn381&180&&&&\\
2 &1.93&0.07&2367&472&10.0&2565&\phn34&&&&\\
3 &1.31&0.10&2870&533&\phn5.1&\phn696&\phn30&&&&\\
4 &0.94&0.16&\phn216&314&\phn1.8&\phn214&\phn25&&&&\\
5 &0.41&0.24&\phn\phn82&155&\phn1.2&\phn\phn93&132&&&&\\
6 &1.97&0.10&2026&399&\phn5.7&1286&\phn16&&&&\\
7 &0.88&0.19&\phn355&364&\phn2.0&\phn180&\phn26&&&&\\
8 &1.07&0.16&\phn847&368&\phn1.6&\phn165&\phn13&&&&\\
9 &1.11&0.17&\phn292&374&\phn1.9&\phn186&\phn13&&&&\\
\enddata
\tablecomments{Same as Table~\ref{table:trunk}, but for CO leaves. Information of infrared source associations to the CO leaves is included in the final column (see Section~\ref{sec:resultYSO}). The full catalog is available as the online material.}
\end{deluxetable}

\begin{figure}[htbp]
\centering
\includegraphics[width=160mm]{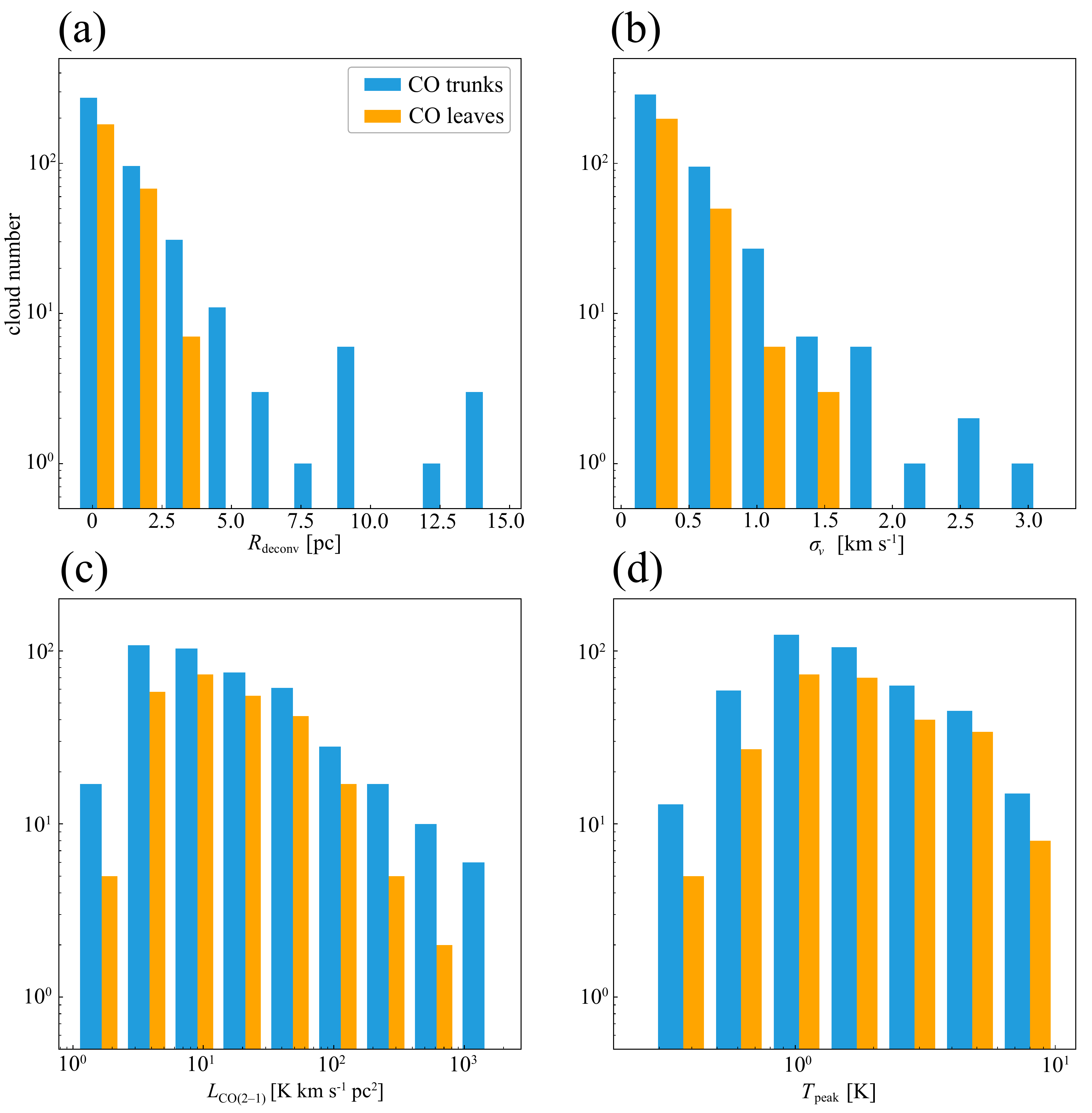}
\caption{Histograms of the physical properties of the CO trunks (cyan) and leaves (orange) in the SMC northem region. Panels (a), (b), (c), and (d) show the deconvolved radius $R_{\rm deconv}$, the velocity dispersion $\sigma_{v}$, the luminosity $L_{\rm CO(2-1)}$, and the peak brightness temperature $T_{\rm peak}$, respectively.}
\label{fig:hist_trunk_leaf}
\end{figure}

Figure~\ref{fig:hist_trunk_leaf} shows histograms of the $R_{\rm deconv}$, $\sigma_{v}$, $L_{\rm CO(2-1)}$, and $T_{\rm peak}$ of the CO trunks and leaves. The total number of luminous large structures is not very large with respect to the full population. The most CO luminous source ($L_{\rm CO(2-1)}$ $\sim$2500\,K\,km\,s$^{-1}$\,pc$^{2}$) is the northern filametnary complex in N66 (see also \citealt{Naslim21}), as shown at the upper left side of Figure~\ref{fig:dendrozoom}(a). For the smaller structure, the CO trunks and leaves seem to exhibit relatively similar properties as a whole.

The physical quantities described above are purely determined from the observational data. Although additional assumptions are needed, we calculated the following properties to further characterize the identified CO sources. We derived the virial mass, $M_{\rm vir} = 1040\sigma_{v}^{2}R_{\rm deconv}$ \citep{Solomon87} assuming the density profile of $\rho \propto r^{-1}$, ignoring the effect of external pressure and magnetic field. 
The peak-integrated intensity was used to calculate the H$_{2}$ column density ($N_{\rm H_2}$) with the assumptions of a CO-to-H$_2$ conversion factor, $X_{\rm CO}$ = 7.5 $\times$10$^{20}$\,cm$^{-2}$\,(K\,km\,s$^{-1}$)$^{-1}$ (\citealt{Muraoka17}) in the SMC and an intensity ratio of CO($J$ = 2--1)/CO($J$ = 1--0), $R_{2-1/1-0}$ of 0.9 (\citealt{Bolatto03}; Paper~I). 
Note that the $X_{\rm CO}$ factor in the low-metallicity SMC environment is not as tightly constrained as the Galactic value. Based on the recent measurement in the literature, the mass determination accuracy is presumably a factor of two or three at best (see also the discussion and our independent estimation using the current CO data set in Section~\ref{D:Xco}). 
$M_{\rm CO}$ is the total gas mass integrated over the regions inside the lowest contour level of the identified structure. 
We estimated the average H$_2$ number density using the following equation: $n_{\rm H_2} = 3M_{\rm CO}/4\pi \mu m_{\rm H} R_{\rm doconv}^3$, where $\mu$ is the mean molecular weight per hydrogen (2.7), and $m_{\rm H}$ is the H atom mass.  

We further explain the relation among the cloud properties, such as the size-linewidth relation, and the cloud mass function in Sections~\ref{sec:r-v}, \ref{D:r-v} and \ref{sec:COmf}. We also performed cross-matching analyses with the CO trunks and infrared young stellar sources in Section~\ref{sec:resultYSO}.


\subsection{Size--Linewidth Relation} \label{sec:r-v}
\begin{figure}[htbp]
\centering
\includegraphics[width=180mm]{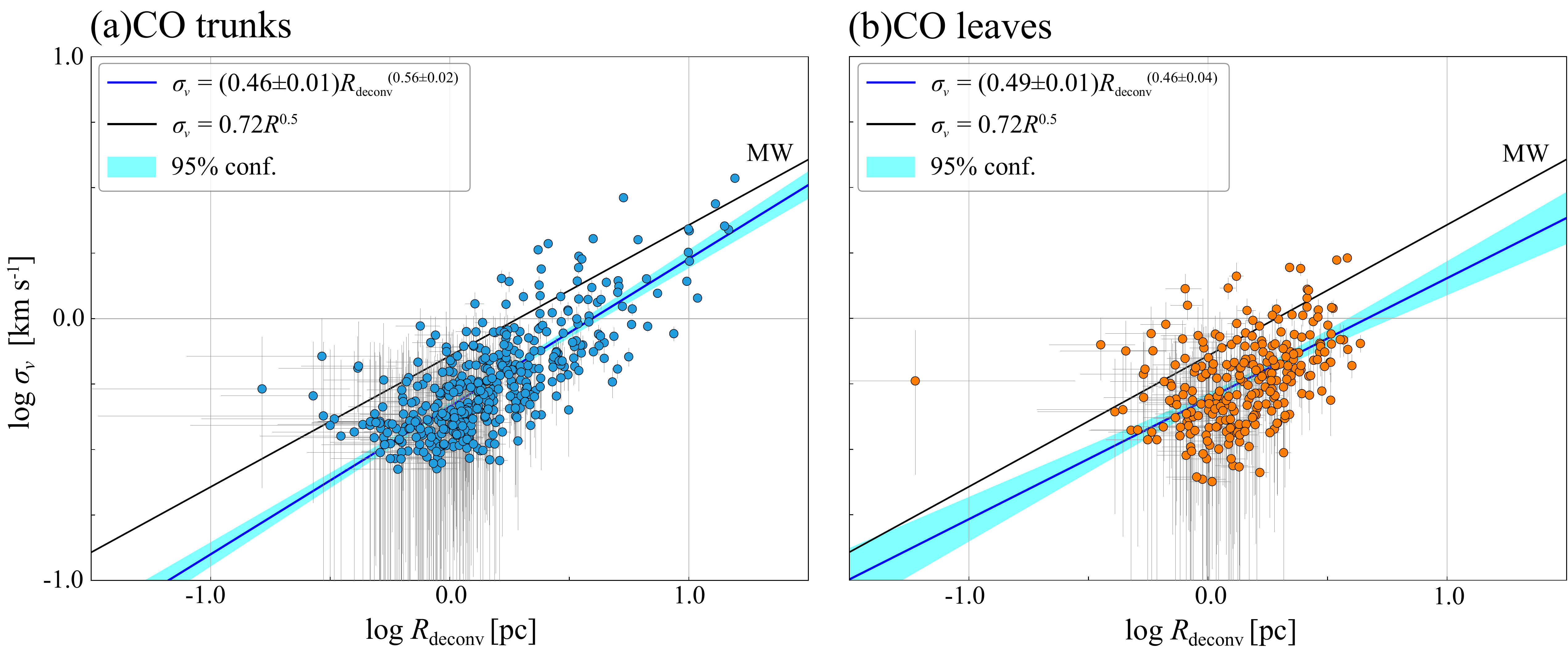}
\caption{Size($R_{\rm deconv}$)--linewidth($\sigma_{v}$) plots of the CO trunks and leaves. Cyan and orange circles denote CO trunks and leaves, respectively. The blue line shows the best-fit functions with the ODR fitting. The constants and power-low indices are shown in the figure legend. The cyan hatch represents the 95\% confidence interval for linear regression. The black line denotes the size--linewidth relation derived in CO($J$ = 1--0) observations of MW molecular clouds \citep{Solomon87}.}
\label{fig:r-v}
\end{figure}

Large-scale molecular cloud surveys found the famous scaling relation between the molecular cloud radius $R$ in pc units and the velocity dispersion $\sigma_{v}$: that is, $\sigma_{v} \approx 0.72R^{0.5}~{\rm km\,s^{-1}}$ \citep[e.g.,][]{Larson81,Solomon87,Heyer01}. This relation is established over a wide spatial range from $\sim$1\,pc to several hundred pc. The sizes (radii) of our CO cloud sample identified as trunks range from $\sim$1\,pc to a few dozen pc, which allows us to test whether a similar relation to the MW is also valid in the SMC over an order of magnitude. Figure~\ref{fig:r-v} shows the $\sigma_{v}$--$R_{\rm deconv}$ plot of the CO trunks and leaves: $\sigma_{v}$ becomes larger as $R_{\rm deconv}$ increases. 

We performed an orthogonal distance regression (ODR) fitting, taking into account the errors in both axes \citep[\texttt{scipy.odr};][]{Virtanen20}, to determine the intercept and slope of $\sigma_{v} = \alpha_{0} R^{-\beta}$. As shown in Figure~\ref{fig:r-v}, the best-fit values are ($\alpha_0$, $\beta$) = (0.46$\pm$0.01, 0.56$\pm$0.02) for the CO trunks and ($\alpha_0$, $\beta$) = (0.49$\pm$0.01, 0.46$\pm$0.04) for the CO leaves. The fitted intercepts are $\sim$0.2 lower than that in the MW standard relation, while the power-low index is comparable to that of the MW. The recent CO(2--1) SMC survey at a 9\,pc resolution also reproduced a similar trend \citep{Saldano23}. We discuss the implications of the size--linewidth relation in Section~\ref{D:r-v}.

\subsection{Cross-matching Analysis between CO and Infrared Sources} \label{sec:resultYSO} 

We investigated whether the CO trunks have known infrared sources with their categories of (1) Spitzer + Herschel young stellar object (YSO) candidates and (2) not necessarily categorized as YSO, but infrared point sources discovered by Spitzer. Based on a better infrared position accuracy than the beam size of the ACA, we regarded a CO trunk as an associated source if there was at least a single infrared source within its cloud boundary.

\begin{figure}[htbp]
\centering
\includegraphics[width=140mm]{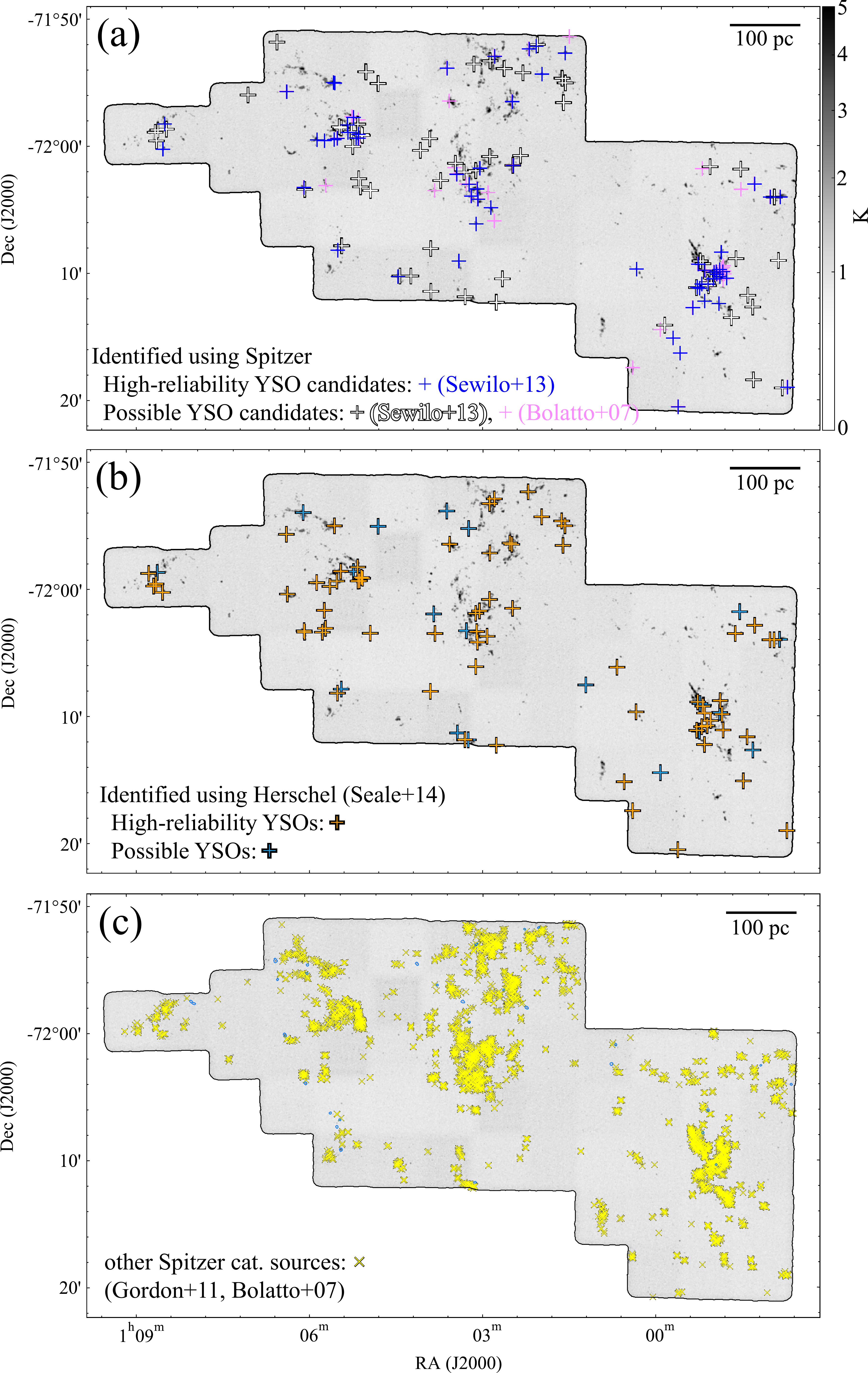}
\caption{Distributions of infrared sources on the CO($J$ = 2–1) peak brightness temperature map of the SMC northern region.
(a) Blue crosses denote high-reliability YSO candidates, and white and pink crosses denote possible YSO candidates \citep{Bolatto07, Sewilo13}. 
Panel (b): Same as panel (a), but the orange and cyan crosses denote the high-reliable and possible YSOs, respectively, identified by \cite{Seale14} using Herschel data. Panel (c): Same as panel (a), but the yellow crosses denote the position of the other Spitzer catalog sources \citep{Bolatto07,Gordon11} associated with the CO clouds. Blue contours highlight the starless cloud candidates (see the text).}
\label{fig:dendroYSO}
\end{figure}

\cite{Gordon11} obtained a comprehensive point-source catalog from the Spitzer Space Telescope Surveying the Agents of Galaxy Evolution in the Tidally Stripped, Low Metallicity Small Magellanic Cloud (SAGE--SMC) Legacy Program. The SAGE-SMC IRAC (InfraRed Array Camera) Single Frame + Mosaic Photometry Catalog has an angular resolution of $\sim$2$\arcsec$ at IRAC bands (3.5/4.5/5.8/8.0\,$\mu$m) with a pointing accuracy of $\sim$0\farcs3 (see the documentation by \citealt{Meade14}\footnote{https://irsa.ipac.caltech.edu/data/SPITZER/SAGE-SMC/docs/sage-smc\_delivery\_nov09.pdf}), which is sufficiently high to be compared with the ACA CO map at $\sim$7$\arcsec$ resolution.The Spitzer/SAGE-SMC point-source catalog includes not only YSOs, but also many normal stars, evolved stars, and background galaxies (see e.g., \citealt{Boyer11}). Several studies identified and characterized the young population based on the color-magnitude-diagram (CMD), spectral energy distribution (SED) modeling by combining data from other wavelengths. 
There is a list of 4927 objects in the SMC that has at least two or more band identifications as point sources in the IRAC(3.5/4.5/5.8/8.0\,$\mu$m) or MIPS 24\,$\mu$m detectors and that satisfy certain CMD criteria to exclude contamination from background galaxies and evolved stars (see Section~4.1 in \citealt{Sewilo13}).
\cite{Sewilo13} identified 742 high-reliability YSO candidates across the SMC based on the CMD color-magnitude cuts, image inspection, SED fitting, and a CMD score (a measure of confidence that a source is not a non-YSO contaminant, based on its position in CMDs used for the initial source selection). Out of these, 452 candidates are well characterized by YSO SED models \citep{Robitaille06}. Within the ACA observed field, the total number of the Spitzer-based YSO candidates is 254; they are plotted in Figure~\ref{fig:dendroYSO}(a). \cite{Seale14} extended the YSO search to longer wavelengths based on the HERschel Inventory of the Agents of Galaxy Evolution (HERITAGE) data \citep{Meixner13}.
Figure~\ref{fig:dendroYSO}(b) shows the identified candidates, which are the high-reliability + possible YSOs in the \cite{Seale14} catalog. In the ACA observed field, there are 25 YSO candidates that were not cataloged in the Spitzer mid-infrared studies above, indicating that they are likely younger. We call them (1) Spitzer + Herschel YSO candidate list and investigate whether they are contained within the lowest contours of the CO trunks.

The positions of the Spitzer + Herschel YSO candidates show a good spatial correlation with the CO cloud distributions, indicating that they are true YSOs enveloped in their natal molecular material. However, due to the CMD selection criteria, these highly reliable YSO samples are mostly biased toward high- and intermediate-mass objects \citep{Sewilo13}. In addition, the angular resolution of the previous CO survey (e.g., $\sim$160\arcsec; \citealt{Mizuno01}) was two orders of magnitude coarser than that of Spitzer, making it impossible to accurately investigate whether the IRAC point sources are spatially correlated with molecular clouds. Our analysis of the CO cloud association with the full IRAC/MIPS catalog potentially allows us to search for additional YSO candidate samples. 
We conducted a cross-matching between the SAGE-SMC catalog sources and our CO data and found that 336 CO trunks were associated, while the remaining 90 entities did not match the catalog. Additionally, we compared with the S$^3$MC (Spitzer Survey of the Small Magellanic Cloud) catalog \citep{Bolatto07,Simon07}, which is based on a deeper survey than SAGE-SMC. The combined SAGE-SMC and S$^3$MC source lists are collectively referred to as $``$other Spitzer catalog sources".
We first checked whether the list (1) is in the CO clouds, and if it was not, (2) we investigated whether the other Spitzer source list was attached within them. For display purposes, we only plotted the (2) sources with a CO detection within the lowest contours of the trunks (Figure~\ref{fig:dendroYSO}(c)).

If there is no other Spitzer catalog source in the CO trunks, it is regarded as a starless cloud candidate, highlighted in blue contours in Figure~\ref{fig:dendroYSO}(c). 
It should be noted that according to the current criteria for infrared catalog extraction, these candidates are still considered to be in a purely starless phase. Upon our visual inspection of the IRAC maps, some sources with extended emission also have local peaks that appear to be associated with CO clouds. Furthermore, even in sources that are completely dark in the Spitzer survey, high-resolution molecular gas studies have sometimes discovered molecular outflow as a strong indicator of protostar formation in infrared-quiescent regions in the MW \citep[e.g.,][]{Tan16} and the LMC \citep[e.g.,][]{Tokuda19,Tokuda22}. 
The James Webb Space Telescope (JWST) will enable us to detect such faint sources that are missed with Spitzer. However, because these sources are low-mass sources or are in an early stage of high-mass star formation, we believe that their feedback effect on the parental cloud itself is negligible on a large scale, and it is not deeply explored in this work.

Column (21) of Table~\ref{table:trunk} denotes the cross-matched results. The 426 CO trunks in total (see Section~\ref{sec:dendro}) can be divided into three categories: 94 Spitzer + Herschel YSO sources, 303 Spitzer catalog sources, and 29 starless cloud candidates. To facilitate the comparison among the categories, the following analysis excludes the CO trunks with CO leaves, i.e., complex, large structures. Table~\ref{table:CO/IR} summarizes the typical (median) properties of the single CO trunks of each category. The resulting number of Spitzer + Herschel sources, other Spitzer catalog sources, and starless cloud candidates are 57, 275, and 29, respectively.
We performed a Kolmogorov-Smirnov (KS) test to determine whether the physical properties belonged to different populations. The $p$-values for the Spitzer + Herschel YSO, other Spitzer catalog, and starless candidate source properties are all below 0.05, except for $n_{H_2}$. Nevertheless, we argue that all the samples belong to distinct populations.

\begin{deluxetable}
{lcccccccccc}
\tabletypesize{\small}
\tablecaption{The median properties of the three categories of the single CO trunks in the SMC North \label{table:CO/IR}}
\tablewidth{0pt}
\tablehead{
\colhead{Category} & \colhead{Number} & \colhead{$\sigma_v$} & \colhead{$\pm\sigma_v$} & \colhead{$L_{\rm CO(2\text{--}1)}$} & \colhead{$\pm L_{\rm CO(2\text{--}1)}$} & \colhead{$R_{\rm deconv}$} & \colhead{$\pm R_{\rm deconv}$} \\
&&\colhead{[km s$^{-1}$]} & \colhead{[km s$^{-1}$]} & \colhead{[K km s$^{-1}$ pc$^{2}$]} & \colhead{[K km s$^{-1}$ pc$^{2}$]} & \colhead{[pc]} & \colhead{[pc]}}
\startdata
Spitzer + Herschel YSO candidate       	& \phn57 & 0.66 & 0.25 & 31.9 	 & 47.3 	& 1.57 & 0.80 \\
other Spitzer catalog source 	& 275 	 & 0.46 & 0.17 & \phn7.2 	 & 20.9 & 1.14 & 0.66 \\
Starless cloud candidate 	& \phn29 	 & 0.40 & 0.13 & \phn4.2 & \phn4.1 & 0.91 & 0.38 \\ \hline
\colhead{Category} & \colhead{$N_{\rm H_2}$} & \colhead{$\pm N_{\rm H_2}$} & \colhead{$M_{\rm CO}$} & \colhead{$\pm M_{\rm CO}$} & \colhead{$n_{\rm H_2}$} & \colhead{$\pm n_{\rm H_2}$} & \colhead{total $M_{\rm CO}$}\\
&\colhead{[10$^{21}$ cm$^{-2}$]} & \colhead{[10$^{21}$ cm$^{-2}$]} & \colhead{[$M_{\odot}$]} & \colhead{[$M_{\odot}$]} & \colhead{[10$^{2}$ cm$^{-3}$]} & \colhead{[10$^{2}$ cm$^{-3}$]} & \colhead{[10$^{4}$ $M_{\odot}$]}\\ \hline
Spitzer + Herschel YSO candidate       	& 3.9 & 2.5 & 574 & 853 	& 14 & 23 	& 4.8 \\
other Spitzer catalog source 	& 1.1 & 1.1 & 130 & 377 	& 10 & 49 	& 7.1 \\
Starless cloud candidate   	& 0.7 & 0.4 & \phn75 & \phn73 & 10 & 29 	& 0.3 \\
\enddata
\tablecomments{$\pm$ denotes the standard deviation of each physical property. We adapted $X_{\rm CO}$ = 7.5 $\times$10$^{20}$\,cm$^{-2}$\,(K\,km\,s$^{-1}$)$^{-1}$ to obtain column density ($N_{\rm H_2}$), cloud mass ($M_{\rm CO}$), and number density ($n_{\rm H_2}$). Total $M_{\rm CO}$ is the sum of $M_{\rm CO}$ in each category.}
\end{deluxetable}

\begin{figure}[htbp]
\centering
\includegraphics[width=190mm]{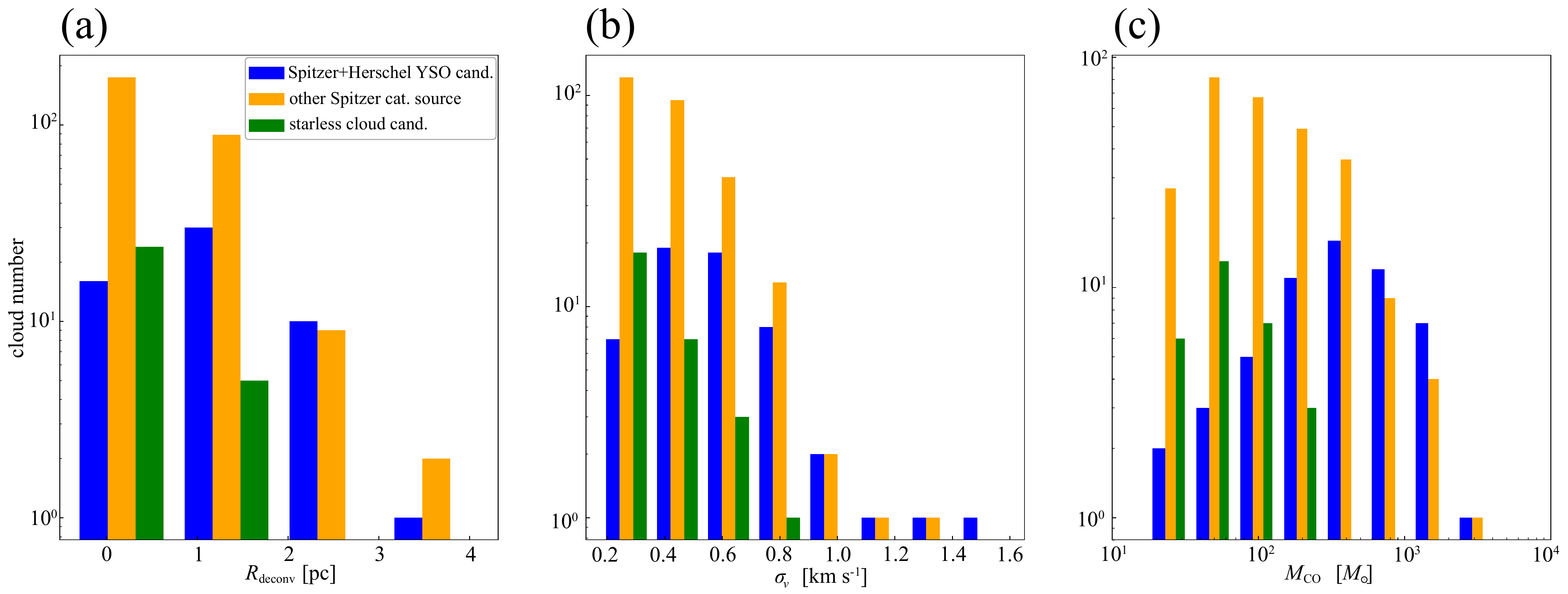}
\caption{Histograms of the physical properties of the single CO trunks in the SMC northern region. Panels (a), (b), and (c) show the deconvolved radius $R_{\rm deconv}$, the velocity dispersion $\sigma_{v}$, and the CO luminosity-based mass $M_{\rm CO}$, respectively. Blue, orange, and green bars denote the number of single CO trunks with Spitzer + Herschel YSO, other Spitzer catalog sources, and starless candidate sources, respectively.} 
\label{fig:hist}
\end{figure}

The total number and $M_{\rm CO}$ of the starless sources correspond to $\sim$8\% and $\sim$2\%, respectively, with respect to the total population (see Table~\ref{table:CO/IR}). $\sigma_{v}$ and $M_{\rm CO}$ appear to be larger as the star formation activity becomes energetic. The typical $\sigma_v$ in the Spitzer + Herschel YSO sources is indeed larger than the value that we expect from the global size--linewidth relation (Section~\ref{sec:r-v}) at the $R_{\rm deconv}$. 
Figure~\ref{fig:hist} shows the comparison histogram of the physical properties. The general trend is that large physical quantities are in the two categories: Spitzer + Herschel YSO candidate, and other Spitzer catalog source.

Single-dish Galactic and ALMA LMC studies also obtained a higher velocity dispersion and larger radius/mass at star-forming clouds \citep{Kawamura98,Ikeda09,Nayak16,Naslim18}. They discussed that feedback from protostellar objects, such as high-radiation pressure of shocks and molecular outflow/jets, enhances the linewidth. The increase in $M_{\rm CO}$ suggests that there is mass accumulation during the star and/or cloud formation phase. 
The possible mass-supply sources are CO-dark-H$_2$ and/or H$\;${\sc i} gas around the CO clouds, as suggested in LMC studies \cite[e.g,][]{Fukui19,Tokuda19,Tokuda22}. 
According to some theoretical studies, atomic gas is a more important reservoir to promote star formation in a lower metallicity environment \citep[e.g.,][]{Krumholz12,Fukushima20}. 

Interestingly, we found many compact CO clouds whose location is relatively isolated from the larger clouds in the field (see also Paper~I). 
In these clouds, massive YSOs do exist at some of the isolated compact clouds, and they could be suitable targets in which to explore the initial condition of high-mass star formation because the relatively simple configuration provides an easier way than typical molecular cloud complexes that harbor well-developed H$\;${\sc ii} regions and/or supernova remnants. Extragalactic studies are more appropriate for discovering such an object, and recent ALMA observations have been studying similar targets in the LMC \citep{Harada19}. Follow-up ALMA 12 m array observations in the SMC are desired to further understand the nature of these isolated clouds and star formation therein.

\section{Discussions} \label{sec:dis}

\subsection{Meaning of the Size--Linewidth Relation}\label{D:r-v}

\cite{Bolatto08} already described that the velocity dispersions are smaller for clouds with the same sizes compared with the MW relation by a factor of two in lower-metallicity targets of their sample (see also \citealt{Saldano23}). They probably overestimated the cloud sizes due to the larger beam size of $\sim$10\,pc. Our ACA observations show that the size--linewidth relation is closer to that in the MW than the \cite{Bolatto08} result, possibly thanks to the improved spatial resolution. However, we still see a departure from the MW relation toward the lower side in velocity dispersions with a factor of $\sim$1.5. \cite{Bolatto08} discussed two possibilities for this trend: (1) the column density is lower than the MW under the condition of virial equilibrium, or (2) the turbulent motion is not strong enough to stabilize the core, and the clouds are supposed to be unstable against the freefall collapse. In the former case, the column density is proportional to the square of linewidth, i.e., our finding of $\sim$1.5 times lower velocity dispersion in the SMC predicts a factor of $\sim$2 lower column density. 
The second idea is highly unlikely in the MW because statistical counting methods using a large number of starless cores with respect to star-forming cores tell us that the lifetime of dense objects until protostar formation is generally longer than the freefall time \citep{Onishi02,Ward-Thompson07} unless their central density exceeds $\sim$10$^{6}$\,cm$^{-3}$ \citep{Tokuda20}. The derived density range of the CO clouds is on the order of 10$^2$\,cm$^{-3}$~(Table~\ref{table:CO/IR}), and although it might be slightly higher, around 10$^4$\,cm$^{-3}$, as suggested by early studies (\citealt{Muraoka17}; Paper~I), it is unlikely that all of these less-dense clouds are undergoing freefall collapse. We note that the above-mentioned dense core surveys in the MW \citep[e.g.,][]{Ward-Thompson07} constrained the starless cloud densities using multiple molecular lines with a higher spatial resolution as well as independent measurements, such as millimeter/submillimeter continuum observations. Our current SMC study has a single CO line with a lower spatial resolution, and thus it is likely that the uncertainty of the density estimation is quite large compared to the above MW surveys. Moreover, it is difficult to prove whether the starless sources are truly $``$starless$"$ down to a low-mass star regime in the SMC as well.
These observational limitations should be overcome to constrain the timescale of starless molecular clouds more precisely and to further explore the implications of the size--linewidth relations by future studies.

\subsection{The CO Cloud Mass Spectrum} \label{sec:COmf}
The frequency distribution of the mass of the molecular cloud is presented as $dN/dM \propto M^{-\alpha}$ or in the cumulative form, $N(>M) \propto M^{-(\alpha-1)}$. This observed quantity is relevant to the fundamental problem of star formation, how molecular clouds transform into stars, i.e., the origin of the initial mass function. From a galactic perspective, an ensemble of formation and destruction processes of molecular clouds likely determines the cloud mass function \citep{Inutsuka15,Kobayashi17}. Although various CO surveys have been revealed, the cloud population along the MW Galactic plane and nearby galaxies, weak CO emission in metal-poor environments, such as the SMC, makes it difficult for us to accumulate a sufficient sample to know the cloud mass function. \cite{Saldano23} obtained a sufficient number ($>$100) of CO clouds in the SMC for the first time and derived the mass spectrum. Our ACA observations still give us further constraints down to the low-mass regime where the CO emission is not clearly visible in the previous single-dish measurement.

\begin{figure}[htbp]
\centering
\includegraphics[width=180mm]{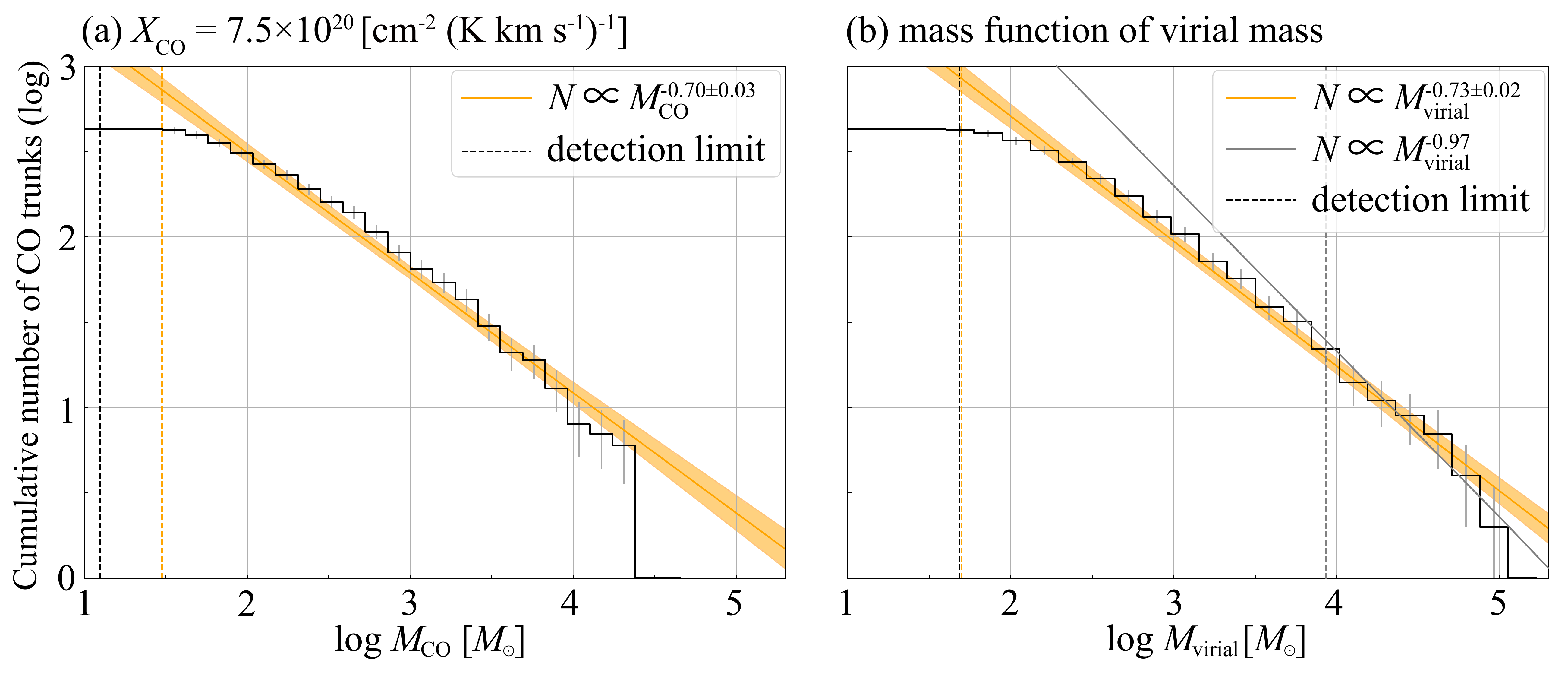}
\caption{Cumulative $M_{\rm CO}$ spectra of the CO trunks using the luminosity-based mass (a) and virial mass (b). 
Mass ranges larger than the vertical dashed orange line were used for the ODR fitting, and the best-fit power-law function for each fitting is shown in the figure legend. The orange hatch denotes the 95\% confidence interval for the fitting. Mass detection limits are shown by dashed black lines in each panel. The solid gray line in panel (b) indicates the fitted slope with a mass range of $>$8.6 $\times$10$^{3}$\,$M_{\odot}$. 
}
\label{fig:Mfunc}
\end{figure}

We use the trunks, which are spatial or velocity-isolated components defined by a low-level contour and are assumed to be less sensitive against the \texttt{astrodendro} parameters. 
Figure~\ref{fig:Mfunc} represents the cloud mass spectra of the luminosity and virial mass with the cumulative form. The features are very similar between the two spectra, except for the presence of massive clouds in the virial mass plot. 
We performed the ODR fitting to the mass spectra and reasonably characterized them by a single power law across two of three orders of magnitudes in the mass range with an exponent of $\sim$0.7, corresponding to $\alpha$ $\sim$1.7. \cite{Takekoshi17} reported a similar value{, $\alpha$ = 1.76 with their completeness limit of 8 $\times$10$^{3}$\,$M_{\odot}$}, by compiling the 1.1\,mm continuum selected Giant Molecular Clouds (GMCs) across the SMC. 

We compare the derived mass spectrum index, $\sim$1.7 with the previous CO study in the SMC. \cite{Saldano23} reported a steeper power-low index of $\alpha$ = 3.1--3.5 in the same region, N66 + NE, in their paper. The discrepancy is presumably caused by the following three factors. (1) The field coverage of our ACA study is wider than that of the APEX observations \citep{Saldano23}. The molecular clouds in the SMC northern region are more sparsely distributed than in the southwestern region. The limited field coverage with APEX did not capture some of the massive CO clouds. The SW region, where many CO clouds are densely packed into almost the same area as the NE coverage, shows shallower mass spectra. (2) The fitting mass ranges are different from each other. We performed a fitting to the $M_{\rm vir}$ function of our data in the same range as \cite{Saldano23}, $>$8.6 $\times$10$^{3}$\,$M_{\odot}$, and obtained a steeper index, $\alpha$ $\sim$2 (see Figure~\ref{fig:Mfunc}b). (3) The resultant index of the cloud mass function somewhat depends on the observation sensitivity and the decomposition algorithm \citep[e.g.,][]{Pineda09}. Their analysis using CPROPS \citep{Rosolowsky06} extracted local maxima of the CO emission, possibly causing an oversegmentation for larger clouds. It is not necessarily consistent with our trunk-based identification, whose cloud boundaries are well characterized by the lowest contour level. Considering some observational and methodological limitations, $\alpha$ $\sim$1.7 derived by our study in the SMC northern region and/or $\alpha$ $\sim$2 derived by \cite{Saldano23} in the other SMC regions would currently be appropriate values to represent the CO cloud mass function of the galaxy.

We subsequently compare the CO cloud mass spectrum in the SMC with the spectra in the MW and LMC studies at galactic scales.
\cite{Heyer01} and \cite{Fukui08} reported that the indices $\alpha$ of CO cloud spectra are $\sim$1.8 in the MW and LMC. These are consistent with our results. Note that a higher-resolution survey in the LMC by \cite{Wong11} reported a much steeper value, possibly because the larger clouds are resolved into smaller ones, which probably is the same as the third issue in the previous paragraph. Although the CO emission likely cannot trace a large amount of molecular material in the metal-poor environment (e.g., \citealt{Glover12,Fukushima20,Bisbas20}, see also Paper~I), it is still intriguing that the Local Group of galaxies shows a similar behavior in CO cloud mass function. M. I. N. Kobayashi et al. (2023, in preparation) numerically demonstrated that mass functions of cold neutral medium, which eventually evolve into molecular clouds, show a spectrum index of 1.7 and do not largely depend on the metallicity condition with $Z$ = 0.2--1.0\,$Z_{\odot}$ after sufficient cooling time under the same conversing H$\;${\sc i} flow setting. It will be important in the future to develop a theory and/or numerical models of molecular cloud formation that take into account the CO abundance and compare these models with observations.

\cite{Inutsuka15} formulated that the exponent of the mass function is determined by the ratio of the formation and destruction timescale ($T_{\rm f}$, and $T_{\rm d}$) of molecular clouds and suggested that the theory explains observed indices of $\alpha$=1.5--2.0 well if $T_{\rm d}$ is longer than $T_{\rm f}$ \citep[see also][]{Kobayashi17}. 
We also remark that there is a mass truncation at $\sim$10$^{4}$--10$^{5}$\,$M_{\odot}$ in the SMC northern spectra. The mass truncation is determined by the total amount of parental material, i.e., H$\;${\sc i} \citep{Kobayashi17}.  
These environments do not harbor many high-mass stars, making superbabble-type H$\;${\sc i} flows, which would be a supply source that might trigger massive GMC formation, and thus can provide the mass truncation in quiescent interarm regions in the MW and M51 \citep{Kobayashi17,Kobayashi18}. Because \cite{Saldano23} also argued that low-mass clouds are dominant in the SMC, additional interferometric studies such as our ACA observations toward other regions would provide further insight into the CO cloud mass function and its regional dependence in the low-metallicity SMC.

\section{Summary} \label{sec:sum}

The CO($J$ = 2--1) ACA survey in the SMC northern region with a field coverage of $\sim$0.26\,deg$^2$ is a powerful map based on which the CO cloud population and properties can be comprehensively understood. Its size scale rages from $\sim$1 pc to a few dozen pc. Our analysis and the obtained implications are summarised as follows:   

\begin{enumerate}
\item Using the \texttt{astrodendro} package, we have decomposed the observed CO clouds into 426 spatially and velocity-isolated components surrounded by a low-level isosurface contour (trunks) and 257 smaller internal structures (leaves). Out of all of the identified structures, $\sim$85\% of the trunks do not have internal leaf substructures (single CO trunks), indicating that many compact/isolated clouds exist throughout the observed field. Based on the cross-matching analysis with the know infrared sources that are cataloged based on Spitzer and Herschel studies, a large fraction of (more than 90\%) of the single CO trunks harbors infrared sources are most likely YSOs.

\item The size--linewidth relation for CO clouds (trunks and leaves) tends to show a smaller linewidth as a whole than that in the MW with a factor of $\sim$1.5. Although an independent single-dish CO study \citep{Saldano23} also confirmed this trend, our parsec-size beam size measurement further constrains this down to small radii of the CO clouds in the unbiased higher-resolution study. One possible interpretation of the lower velocity dispersion is that the column density is a factor of $\sim$2 lower than the densities in the MW clouds, assuming that the cloud is well supported against the free-fall collapse. 

\item The CO-luminosity-based mass and virial mass spectra of the CO trunk in the cumulative form follow power-law indices of $\sim-$0.7, corresponding to $dN/dM \propto M^{-1.7}$. The power-law index is similar to the indices from CO surveys of the MW and LMC. Although the CO dark fraction with respect to the total molecular material in the SMC is likely higher than in the two galaxies, the striking similarity of the CO cloud mass function may be one of the milestones for understanding molecular cloud formation and their metallicity (in)dependence from a theoretical perspective.

\end{enumerate}

\acknowledgments

We would like to thank the anonymous referee for useful comments that improved the manuscript. This paper makes use of the following ALMA data: ADS/ JAO. ALMA\#2017.A.00054.S. ALMA is a partnership of ESO (representing its member states), NSF (USA) and NINS (Japan), together with NRC (Canada), MOST and ASIAA (Taiwan), and KASI (Republic of Korea), in cooperation with the Republic of Chile. The Joint ALMA Observatory is operated by ESO, AUI/NRAO, and NAOJ. This work was supported by NAOJ ALMA Scientific Research grant Nos. 2022-22B and Grants-in-Aid for Scientific Research (KAKENHI) of Japan Society for the Promotion of Science (JSPS; grant Nos. JP18K13582, JP18H05440, JP21K13962, and JP21H00049). The material is based upon work supported by NASA under award number 80GSFC21M0002 (M.S.). The National Radio Astronomy Observatory is a facility of the National Science Foundation operated under cooperative agreement by Associated Universities, Inc. T.W. acknowledges support from collaborative NSF AAG awards 2009849. We thank Dr. Masato I.N. Kobayashi for the discussion on the cloud mass spectrum from theoretical aspects.

\software{CASA (v5.4.0; \citealt{casa2022}), Astropy \citep{Astropy18}, APLpy \citep{Robi12}}

\appendix

\section{Input parameter dependence of the cloud identification}

\begin{figure}[htbp]
\centering
\includegraphics[width=140mm]{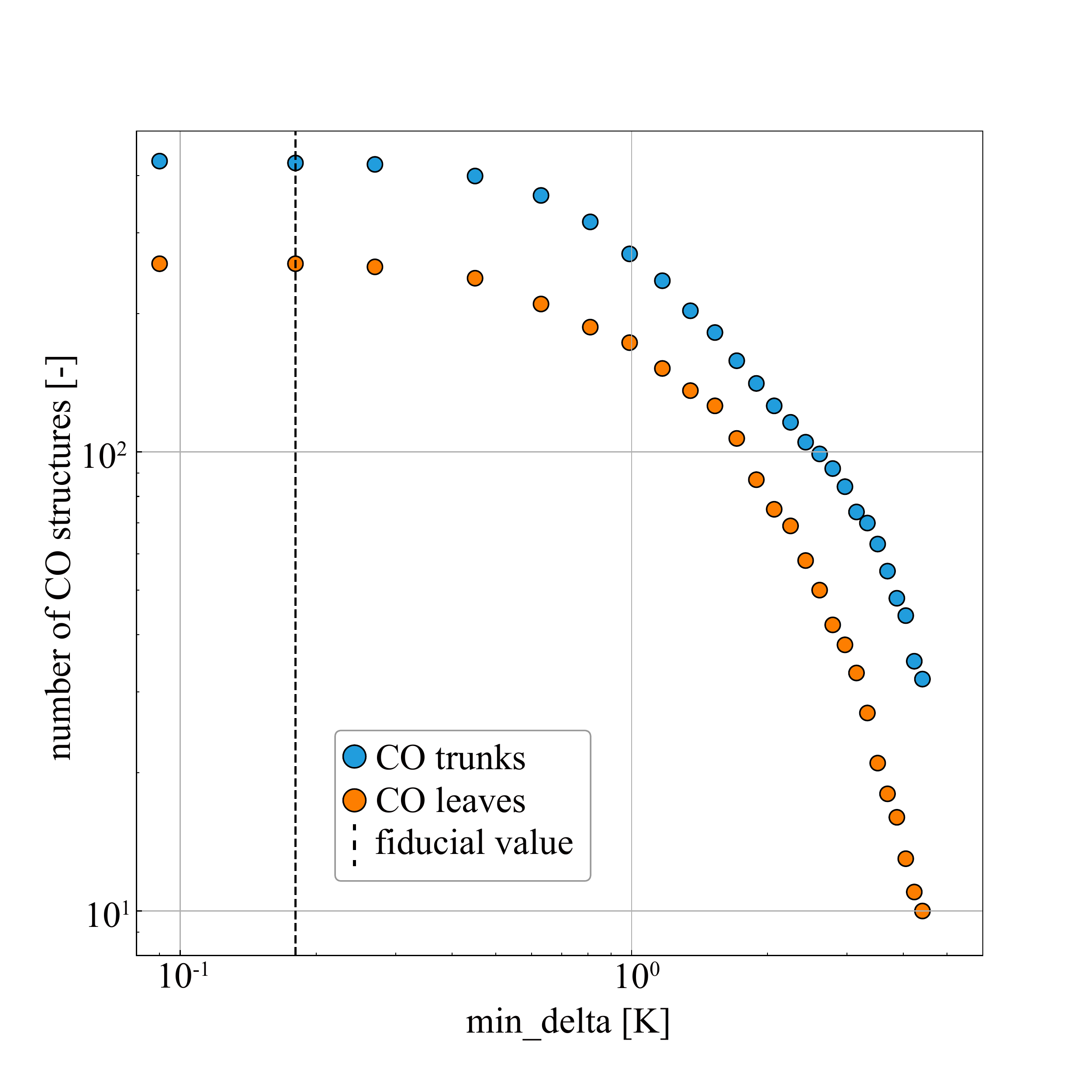}
\caption{Parameter dependence of the astrodendro algorithm on \texttt{min\_delta}. The dashed black lines correspond to the fiducial value of 0.18 K used for our cloud decomposition.}
\label{fig:min_delta}
\end{figure}

As explained in Section~\ref{sec:dendro}, \texttt{min\_delta} is a relatively arbitrary parameter among the astrodendro arguments, especially for spatially compact objects with a well-defined outer boundary. We investigate the \texttt{min\_delta} dependence of the number of identified leaves/trunks (Figure~\ref{fig:min_delta}). The number of identified clouds decreases sharply in the range of \texttt{min\_delta} above 1\,K. This is because the emission does not satisfy the requirement that they should have a difference of more than 1\,K of the brightness temperature within their structure and thus cannot be considered as a single leaf or trunk. 
As a result, in the range with a large \texttt{min\_delta}, only structures with a strong intensity contrast (i.e., the maximum intensity is high as well) survive. 

\section{Virial Mass--CO Luminosity Relation}\label{D:Xco}

\begin{figure}[htbp]
\centering
\includegraphics[width=180mm]{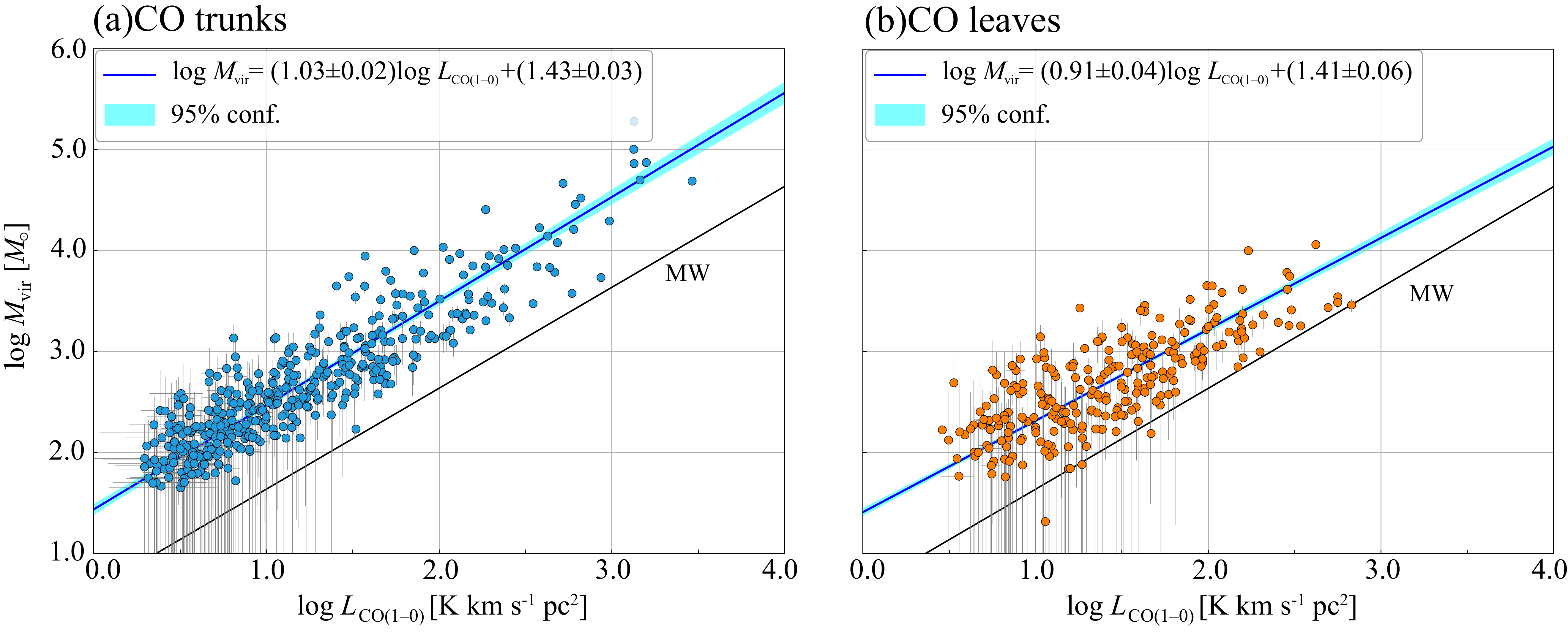}
\caption{$M_{\rm vir}$ vs. equivalent $L_{\rm CO(1-0)}$ plots for the CO trunks (a) and leaves (b). Cyan and orange circles denote CO trunks and leaves, respectively. The blue lines and cyan hatches are the best-fit functions and the 95\% confidence interval for linear regression with the least-squares method. The intercept and slope of the fitting function are shown in the figure legend with errors.}
\label{fig:LcoMvir}
\end{figure}

Figure~\ref{fig:LcoMvir} shows the $L_{\rm CO(1-0)}$ versus $M_{\rm vir}$ relations of the trunk/leaf structures. Note that we converted the observed $L_{\rm CO(2-1)}$ into the equivalent CO($J$ = 1--0) luminosity, $L_{\rm CO(1-0)}$, by adopting an $R_{\rm 2-1/1-0}$ of 0.9 (\citealt{Bolatto03}; Paper~I). The $L_{\rm CO(1-0)}$ and $M_{\rm vir}$ are well correlated with each other as a whole over the range of two orders of magnitude, indicating that the clouds in the observed region are virialized and that the CO luminosity can be a good tracer of mass.

In the extragalactic perspective, the comparison between the two quantities is an almost unique method for estimating the $X_{\rm CO}$ factor with CO measurements alone \citep{Bolatto13} using the following equation: 
\begin{equation}
X_{\rm CO} = \frac{M_{\rm vir}/(m_{\rm H}\mu)}{L_{\rm CO(1-0)}}
\end{equation}
The median values of $X_{\rm CO}$ for the trunks and leaves are 1.3$_{-4.3}^{+0.8}$ $\times$10$^{21}$\,cm$^{-2}$\,(K\,km\,s$^{-1}$)$^{-1}$ and 8.4$_{-1.4}^{+9.6}$ $\times$10$^{20}$\,cm$^{-2}$\,(K\,km\,s$^{-1}$)$^{-1}$, respectively, with the plus and minus signs indicating the first and third quartiles.
As seen in Figure~\ref{fig:LcoMvir}, all of the data points are well above the MW canonical relation, $X_{\rm CO}^{\rm MW} = 2.0\,\times10^{20}\,{\rm cm^{-2}\,(K\,km\,s^{-1})^{-1}}$, indicating that the conversion factor $X_{\rm CO}$ in the SMC northern region is higher than that in the MW. 

We obtained two $X_{\rm CO}$ factors from the identification results for trunks and leaves. We compare the newly derived $X_{\rm CO}$ with that of previous studies in the SMC CO surveys. The NANTEN survey \citep{Mizuno01} with an angular resolution of 45\,pc reported $X_{\rm CO}$ of  $\sim$2.5\,$\times$10$^{21}$\,cm$^{-2}$\,(K\,km\,s$^{-1}$)$^{-1}$ based on the same virial-mass-based method. 
Although their values are close to the value derived in our trunk structure, the lower value is obtained because the fine-beam measurement eliminates the overestimation of the cloud size.

\cite{Bolatto13} cautioned that this virial-mass-based $X_{\rm CO}$ derivation likely overestimates in the weaker CO regime because the total amount of H$_2$ of an extended envelope in CO-free/weak positions is highly ambiguous. They recommended using CO-bright regions reflecting a fairly uniform condition in $X_{\rm CO}$ estimate and implied that their derived $X_{\rm CO}$ value is not significantly different from that in the MW CO clouds. Our derived $X_{\rm CO}$ for the leaves is close to this context because the smaller structure inside the clouds tends to reflect the nature of CO-bright local peaks.
The $X_{\rm CO}$ for the leaves is also consistent with the SEST and ALMA studies toward the N83/N84 regions \citep{Bolatto03,Muraoka17}.
\cite{ONeill22} performed an alternative $X_{\rm CO}$ calibration using the optically thin $^{13}$CO column density estimation in the NGC~602 region and obtained $X_{\rm CO}$ of 3.4 $\times$10$^{20}$\,cm$^{-2}$\,(K\,km\,s$^{-1}$). \cite{Valdivia20} derived virial-mass-based $X_{\rm CO}$, (3--7) $\times$10$^{20}$\,cm$^{-2}$\,(K\,km\,s$^{-1}$) depending on the star formation activities in the Magellanic Bridge (see also \citealt{Kalari20}), where the metallicity is the same as or somewhat lower than that in the SMC main body. The 9\,pc observations through the SMC yielded an etimate $X_{\rm CO}$ based on virial mass and millimeter continuum emission of 2.5 and 6.5 times that of the MW, respectively. The $X_{\rm CO}$ value in the SMC is not tightly constrained with high accuracy, but it seems certain that on average, $X_{\rm CO}$ is several times higher than that in the galaxy. Theoretical studies \citep[e.g.,][]{Feldmann12} indicate that the metallicity dependence of the $X_{\rm CO}$ factor is a power-low function with an exponent of $-$(0.5--0.8).


\bibliography{sample63}{}
\bibliographystyle{aasjournal}

\end{document}